\documentclass[10pt]{article}

\usepackage{amsmath}
\usepackage{amssymb}

\usepackage{graphicx}

\usepackage{cite}

\usepackage{color} 

\usepackage[labelfont=bf,labelsep=period,justification=raggedright]{caption}

\makeatletter
\renewcommand{\@biblabel}[1]{\quad#1.}
\makeatother

\date{}

\begin{document}

% Title must be 150 characters or less
\begin{flushleft}
{\Large
\textbf{Network Algorithmics and the Emergence of Synchronization in Cortical
Models}
}
% Insert Author names, affiliations and corresponding author email.
\\
Andre Nathan, 
Valmir C. Barbosa$^\ast$
\\
\bf Programa de Engenharia de Sistemas e Computa\c c\~ao, COPPE,
Universidade Federal do Rio de Janeiro, Rio de Janeiro, RJ, Brazil
\\
$\ast$ Corresponding author, e-mail: valmir@cos.ufrj.br
\end{flushleft}

% Please keep the abstract between 250 and 300 words
\section*{Abstract}

When brain signals are recorded in an electroencephalogram or some similar
large-scale record of brain activity, oscillatory patterns are typically
observed that are thought to reflect the aggregate electrical activity of the
underlying neuronal ensemble. Although it now seems that such patterns
participate in feedback loops both temporally with the neurons' spikes and
spatially with other brain regions, the mechanisms that might explain the
existence of such loops have remained essentially unknown. Here we present a
theoretical study of these issues on a cortical model we introduced earlier
[Nathan A, Barbosa VC (2010) Network algorithmics and the emergence of the
cortical synaptic-weight distribution. Phys Rev E 81: 021916]. We start with the
definition of two synchronization measures that aim to capture the
synchronization possibilities offered by the model regarding both the overall
spiking activity of the neurons and the spiking activity that causes the
immediate firing of the postsynaptic neurons. We present computational results
on our cortical model, on a model that is random in the Erd\H{o}s-R\'enyi sense,
and on a structurally deterministic model. We have found that the algorithmic
component underlying our cortical model ultimately provides, through the two
synchronization measures, a strong quantitative basis for the emergence of both
types of synchronization in all cases. This, in turn, may explain the rise both
of temporal feedback loops in the neurons' combined electrical activity and of
spatial feedback loops as brain regions that are spatially separated engage in
rhythmic behavior.

% Please keep the Author Summary between 150 and 200 words
% Use first person. PLoS ONE authors please skip this step. 
% Author Summary not valid for PLoS ONE submissions.   
%\section*{Author Summary}

\section*{Introduction}

Current technology allows the recording of electromagnetic brain activity at
several different spatial scales, ranging from invasive recordings that capture
the signals from neuronal ensembles comprising thousands to millions of cells to
those that are noninvasive (like the electroencephalogram) and capture the
signals from large-scale cortical areas \cite{ns06}. Invariably the recorded
signals appear as oscillatory patterns which, depending on the scale at which
the recording is performed, occur in different frequency ranges and provide
distinct interpretive possibilities. At the smallest spatial scales, for
example, the recorded signals occur in the range of a few kilohertz and are
thought to reflect the combined spiking (or firing) activity of the group of
neurons in question. The signals recorded at the largest scales, on the other
hand, occur in the range of a few to $100$ hertz and are thought to reflect the
so-called local field potentials (LFPs), which in turn are purported to reflect
the aggregate effects of all the electrical activity taking place in the
corresponding area \cite{blt12}.

LFPs have acquired great prominence recently, owing mainly to the fact that LFPs
related to different brain areas appear to combine with one another in such
ways as to correlate significantly with the brain's sensory-motor mechanisms and
other, higher-level functions as well (such as memory, attention, and others;
cf.\ \cite{f05,canolty10a,canolty10b} and references therein). The ways in which
LFPs combine seem to involve forms of cross-frequency coupling that cover wide
temporal and spatial ranges and ultimately promote the integration of activities
with different temporal and spatial characteristics \cite{canolty10b}. Decoding
the various forms of LFP coupling may one day hold the key to understanding how
computation and communication take place in the brain.

As it happens, though, many aspects of the nature of LFPs have remained elusive.
In particular, their precise origin and relation to the underlying firing
activity of the neurons seem to depend crucially on the brain region being
considered. However, notwithstanding this indefiniteness that still surrounds
the specifics of LFP emergence and interaction with other LFPs, some of the
crucial points are beginning to be clarified \cite{knbbrc09,blt12}. One of them
is the relation with the combined firing activity of groups of neurons. Although
LFPs seem to derive largely from the accumulation of potential at the neurons'
membranes that eventually leads to neuronal firing, it appears that the firing
patterns themselves have a role to play as well. As a consequence, a picture
that seems to be emerging is that of a feedback loop in which the larger-scale
LFPs both influence and get influenced by the smaller-scale firing patterns.
Another point is more related to the spatial characteristics of how LFPs
interact with one another: It now seems clear that feedback loops also occur
involving the LFPs of distinct brain areas.

While a complete resolution of these issues will certainly require considerable
further research, a detailed understanding will undoubtedly benefit from a clear
picture of firing synchronization at the neuronal level. By synchronization we
mean both the convergence of multiple action potentials onto distinct cells
within a relatively narrow temporal window, and also the firing due to such
potentials within a similar window. As we mentioned above, both phenomena are
closely related to the rise of LFPs, and consequently the rise of the
higher-level functions that LFPs are thought to support. In our view, accounting
for the possibilities the brain affords for the appearance of such synchronized
behavior depends both on the anatomical properties of how neurons interconnect
and on the individual firing behavior of each neuron. Our aim in this paper is
the study of these types of synchronism on a graph-theoretic model of the
cortex.

Our study is preceded by a few others \cite{huerta98,lago-fernandez00,masuda04},
all of which have modeled neuronal behavior as a continuous-time process in
which the relevant signals obey certain differential equations and feed some
measure of synchronous activity. Invariably these studies have modeled the
synchronization of membrane potentials directly, and have for this reason stayed
apart from explicitly modeling most of the relevant details that characterize
neuronal firings, like the so-called local histories of each neuron
\cite{barbour07}. Our approach is to take a different course and focus on the
extent to which certain events at the various neurons can be said to be
synchronized. These events are, in essence, the arrival of action potentials
from other neurons and the eventual creation of new action potentials. These, of
course, are precisely the events that make up membrane potentials, but we have
found in previous studies of related problems that the additional level of
detail pays off by providing unprecedented insight. Examples here are the
emergence of the synaptic-weight distribution, in which case we demonstrated
that experimentally observed distributions \cite{song05} can be reproduced
\cite{nb10}, and a more tractable version \cite{nb11} of the
integrated-information theory for the emergence of consciousness
\cite{balduzzi08}.

Our model is based on what is generally called network algorithmics. In our case
this refers to the combination of a structural component to represent the
anatomical properties of the cortex at the neuronal level, and a
distributed-algorithmic component to represent the traffic of action potentials
among the neurons as they fire. In general, a distributed algorithm is a
collection of sequential procedures, each run by an agent in a distributed
system with provisions for messages to be sent to other agents for
communication \cite{barbosa96}. The distributed algorithm of interest here is
inherently asynchronous, meaning that the behavior of each agent (each neuron)
is purely reactive to the arrival of messages (action potentials), which can in
turn undergo unpredictable delays before delivery. It follows that independent
runs of the same algorithm from the same initial conditions cannot in general be
guaranteed to generate the same sequence of message arrivals at any agent, and
therefore not the same behavior. Lately, this inherent source of
unpredictability has led asynchronous distributed algorithms to be recognized as
a powerful abstraction in the modeling of biological systems \cite{fh07,fhh11},
since they allow the incorporation into the model of small behavioral variations
that might otherwise be simply smoothed out and rendered irrelevant.

An asynchronous distributed algorithm never makes any reference to a
global-time entity. In our model, therefore, every property emerging from
neuronal interactions is the result of strictly local elements. The presence of
such strong asynchronism may seem contradictory with the search for signs of
synchronism. However, we have found that observing the system from the outside
can reveal surprising synchronization possibilities that could never be known
within the scope of individual neurons, provided we uncover the causal
relationship that binds those local elements on a global scale.

Using network algorithmics as the basis of our model makes it substantially more
detailed than those that target membrane potentials as the basic analytical
units, in the sense that a neuron's behavior can now be described as its
reaction to the arrival of a new action potential given its current state.
Clearly, though, there would be room for considerably more detail, even down to
the molecular level. Our approach, therefore, resides in the middle ground
between two extremes and as such is related to what has become known as the
artificial-life approach \cite{f00,l10}, whose ``life as it could be'' metaphor
is close in spirit to our own view. Our approach is also part of a growing
collection of efforts that attempt to tackle eminent problems of neuroscience
from the perspective of graph-based and other interdisciplinary methods
\cite{sporns04,sporns05,achard06,bassett06,he07,honey07,reijneveld07,sporns07,stam07,yhsn08,c10,maenss11}.
As an enterprise that eventually seeks to be able to handle realistically large
problem instances from a graph-theoretic perspective, the present study is
moreover in line with the many others that in the last decade have addressed
the so-called complex networks \cite{bs03,nbw06,bkm09}.

Our model is defined on a strongly connected directed graph $S$, i.e., a
directed graph in which it is possible to reach every node from every other node
through directed paths only. We assume that $S$ has $N$ nodes, each representing
a neuron, so each one can be either excitatory or inhibitory in the proportion
of $4$ to $1$, respectively
\cite{abeles91,ananthanarayanan07,ananthanarayanan09}. In $S$, an edge exists
directed from node $i$ to node $j$ to indicate a synapse from $i$'s axon to one
of $j$'s dendrites, but no edge can connect two inhibitory nodes
\cite{abeles91}. The results we describe in the remainder of the paper for our
cortical model are predicated upon $S$ being drawn from the following
random-graph model on $n\ge N$ nodes in such a way that $S$ is the giant
strongly connected component, GSCC \cite{dorogovtsev01}, of the directed graph
$D$ that is actually drawn. In $D$ a randomly chosen node, say $i$, has
out-degree $k>0$ with probability proportional to
$k^{-1.8}$.\footnote{Originally the choice of this scale-free distribution
\cite{newman05} was inspired by related work that observed such a power law at
a higher level of detail in the cortical architecture
\cite{eguiluz05,vandenheuvel08}, but recent measurements regarding the
distribution of neuronal (undirected) degrees \cite{ms10} seem to support the
choice we made. Specifically, ignoring edge directions in our model leads to an
exponential degree distribution \cite{nb11}, as observed in \cite{ms10}. We also
note that our adoption of a power law to describe the distribution of
out-degrees represents a significant departure from earlier modeling attempts
\cite{abeles91,bs09}, which used randomness in the Erd\H{o}s-R\'{e}nyi sense
\cite{erdos59} directly. We note, moreover, that cortices do exhibit many other
scale-free properties \cite{yhsn08,fkbr09,s11}, including some that evince their
small-world nature \cite{asbs00}, the existence of hubs (nodes with very large
out-degrees), and others \cite{f07,hscgtmh09}.} Following
\cite{kaiser04a,kaiser04b}, each of these $k$ nodes is precisely another
randomly chosen node, say $j$, with probability proportional to $e^{-d_{ij}}$,
where $d_{ij}$ is the Euclidean distance between nodes $i$ and $j$ when all $n$
nodes are placed uniformly at random on a radius-$1$ sphere. We have observed
this policy for edge placement to lead to $N\approx 0.9n$ on average (i.e., we
expect the GSCC of $D$, and hence $S$, to comprise roughly $90\%$ of the $n$
nodes) \cite{nb11}.

\section*{Methods}

We start with the specification of the asynchronous distributed algorithm to be
executed by the nodes of graph $S$. This algorithm prescribes a common procedure
for each node to react to the arrival of a message and to possibly send messages
out as a consequence. Such a procedure runs atomically (i.e., without being
interrupted), so actually implementing the algorithm requires provisions for the
queuing of incoming messages at each node. Each run of the distributed algorithm
can be described formally by specifying how the various message arrivals at the
different nodes relate to one another. As we do this a precise notion of
causality emerges and we use it to define our two synchronization measures.

\subsection*{Asynchronous distributed algorithm}

All nodes in $S$ share the same rest and threshold potentials, denoted by $v^0$
and $v^\mathrm{t}$, respectively, such that $v^0<v^\mathrm{t}$. We use $v_j$ to
represent the potential of node $j$ and $w_{ij}$ to represent the synaptic
weight of the edge directed from $i$ to $j$. At all times, these are constrained
in such a way that $v_j\in[v^0,v^\mathrm{t}]$ and $w_{ij}\in[0,1]$. An
asynchronous distributed algorithm, henceforth referred to as $A$, underlies the
node-potential and synaptic-weight dynamics of our cortical model. At node $j$,
algorithm $A$ is built around the ``firing'' operation which consists of sending
a message to each out-neighbor of $j$ in $S$ and setting $v_j$ to the rest
potential $v^0$. Algorithm $A$ is purely reactive, that is, it only acts at any
node when the node receives a message. Evidently there must be exceptions to
this purely reactive character, since at least one node must send messages out
spontaneously to start the algorithm. We call such nodes initiators and let
$m\le N$ be the number of initiators in a given run of algorithm $A$. When node
$j$ acts as an initiator (this happens at most once in a run for that node), it
simply fires with probability $1$. When it reacts to the reception of a message,
say from node $i$, node $j$ performs the following steps:

\begin{enumerate}
\item If $i$ is excitatory, then set $v_j$ to $\min\{v^\mathrm{t},v_j+w_{ij}\}$.
If it is inhibitory, then set $v_j$ to $\max\{v^0,v_j-w_{ij}\}$.
\item Fire with probability $(v_j-v^0)/(v^\mathrm{t}-v^0)$.
\item If firing did occur during step~2, then set $w_{ij}$ to
$\min\{1,w_{ij}+\delta\}$. If it did not occur but the previous message received
by node $j$ from any of its in-neighbors did cause firing to occur, then set
$w_{ij}$ to $(1-\alpha)w_{ij}$.
\end{enumerate}

The parameters $\delta$ and $\alpha$ appearing in step~3 are constrained by
$\delta\le\alpha$ \cite{nb10}. They are meant to reproduce, although to a
limited extent, the spike-timing-dependent plasticity principles
\cite{abbott00,song00}, which roughly dictate that $w_{ij}$ is to increase if
firing occurs and decrease otherwise, always taking into account how close in
time the relevant firings by nodes $i$ and $j$ are.\footnote{In Hebbian terms,
learning depends on how the synapses incoming to $j$ compete with one another
for the firing of $j$. Strengthening the synapse from $i$, in particular,
depends on whether the arrival of an action potential from $i$ succeeds in
making $j$ fire, either by itself or combined with other recent arrivals that
did not cause firing \cite{song00}.} Moreover, increases in $w_{ij}$ are to
occur by a fixed amount, decreases by proportion \cite{bi98,bi01,kepecs02}.
Every run of algorithm $A$ is guaranteed to eventually reach a point at which no
more message traffic exists. It is then said to have terminated.

\subsection*{Modeling causality}

The problem of handling synchronization during a run of an asynchronous
distributed algorithm such as algorithm $A$ is akin to that of handling
simultaneity in special relativity. This is so because, in either context,
establishing the notion of simultaneity is dependent upon how signals propagate
and also on locality considerations \cite{h11}. A convenient framework within
which to tackle it is then the event-based formalism originally laid down in
\cite{l78}. An event in the present context is said to have occurred either when
an initiator fires (and thereby sends a message to each of its out-neighbors in
graph $S$) or when a node reacts to the reception of a message by executing
steps~1--3 (which includes the possibility of firing). In either occasion the
event involves the execution of an atomic set of actions by the node in
question.

A run of algorithm $A$ can then be regarded as a set $R$ of events. An event
$r\in R$ is described by the $4$-tuple $r=\langle i,t_i,m_i,M_i\rangle$, where
$i$ identifies the node of $S$ at which event $r$ occurred, $t_i\ge 1$ indicates
that $r$ was the $t_i$th event to occur at node $i$ since the run began, $m_i$
is the message (if any) that triggered the occurrence of the event, and $M_i$ is
the set of messages sent by node $i$ if it fired during the event. Many of the
events in $R$ are implicitly related to one another. We make such a relation
explicit by defining the binary relation $B\subseteq R^2$. Given events $r$ as
above and $r'=\langle j,t_j,m_j,M_j\rangle$, we say that the ordered pair
$(r,r')\in B$ if and only if:
\begin{enumerate}
\item[(a)] Either $i$ and $j$ are the same node and $t_i<t_j$ with no
intervening event between $r$ and $r'$;
\item[(b)] Or $j$ is an out-neighbor of $i$ and $m_j\in M_i$ (i.e., the message
that triggered $r'$ was sent in connection with the occurrence of $r$).
\end{enumerate}

We interpret $(r,r')\in B$ as meaning that $r$ happened at node $i$ immediately
before $r'$. If (a) is the case, then this use of ``before'' seems natural
because both $t_i$ and $t_j$ are relative to the same (local) time basis.
Extending this interpretation to case (b), though apparently less natural, is
still consistent because there is no reference in this case to either $t_i$ or
$t_j$. Closing $B$ under transitivity yields another binary relation on $R$,
here denoted by $B^+$, such that $B\subseteq B^+\subseteq R^2$. This relation
can be interpreted in such a way that $(r,r')\in B^+$ means that $r$ happened
before $r'$ regardless of how close the specific nodes at which they occurred
are to each other in graph $S$. Through $B^+$, therefore, relation $B$ is the
core of a characterization of how the events of run $R$ relate to one another
causally. In particular, if $r$ and $r'$ are such that neither $(r,r')\in B^+$
nor $(r',r)\in B^+$, then the situation is analogous to the space-like
separation of events in special relativity (since neither could the occurrence
of $r$ influence that of $r'$ during run $R$, nor conversely).

\begin{figure}[p]
\begin{center}
\scalebox{1.000}{\includegraphics{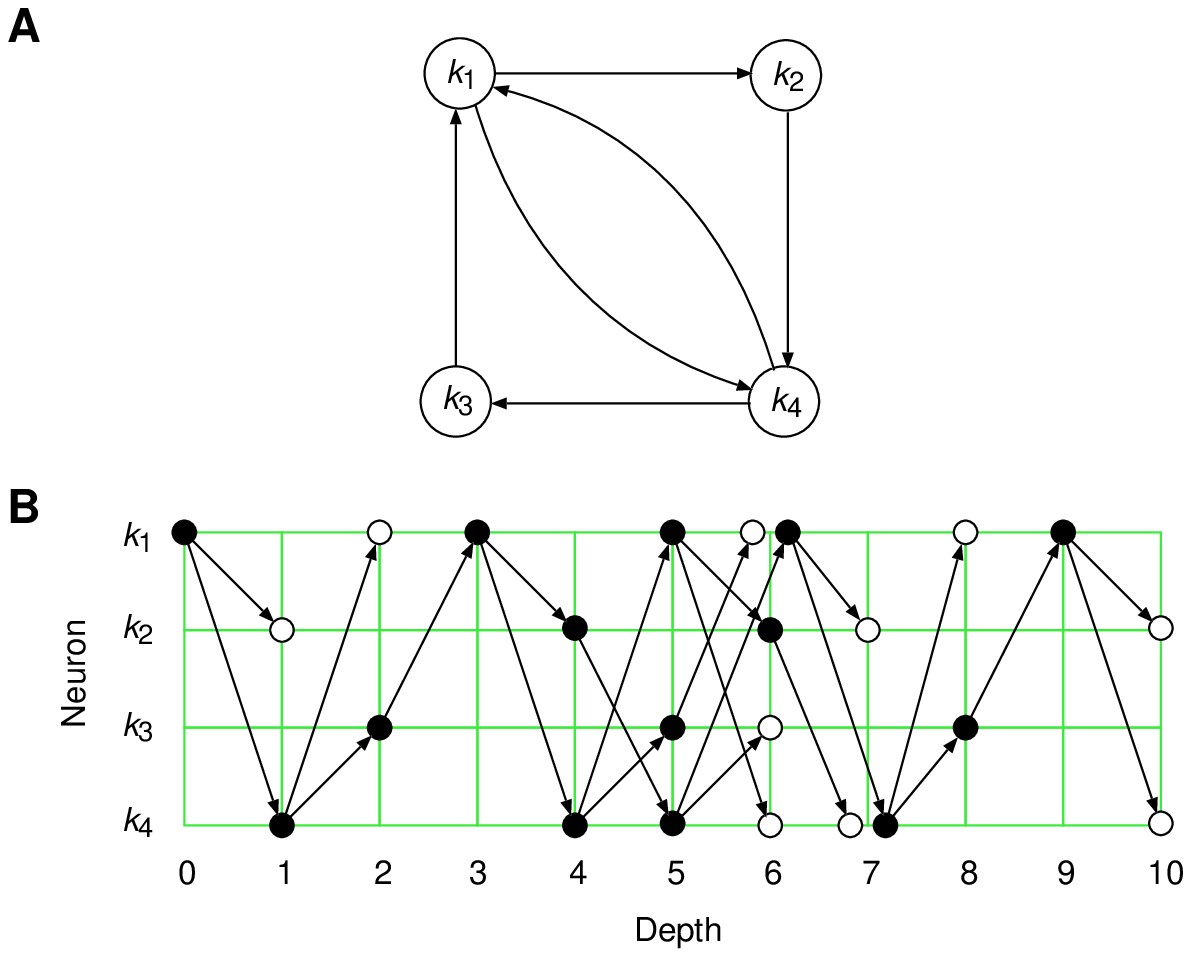}}
\end{center}
\caption{{\bf Directed graphs $S$ and $P$.} A run of algorithm $A$ on graph $S$
(A) started by the single initiator $k_1$ may result in the set of events shown
in panel B as the node set of graph $P$. Events are shown in panel B either as
filled circles, in the case of firing events, or as empty circles, otherwise.
Each event is positioned against a background grid that highlights the node of
$S$ at which it occurred (this is given by the grid's rows) and its depth (the
grid's columns). An edge of $P$ either joins consecutive events occurring at the
same node of $S$ or corresponds to a message. We omit the former from panel B
for clarity. An event's depth is the greatest number of message edges on a
directed path arriving at it. Every event (except the first one at the
initiator) has exactly one incoming message edge. If it is a firing event, then
it also has as many outgoing message edges as the corresponding node in graph
$S$ has out-neighbors. Whenever two or more events occurring at the same node of
$S$ have the same depth, they are shown as close to the correct grid point as
possible.}
\label{fig:graphs}
\end{figure}

Relation $B$ is also instrumental in our characterization of synchronization in
run $R$. The first step is to recognize that it naturally gives rise to a
directed graph, call it $P$,  whose node set is $R$ (the set of events) and
whose edge set is $B$. This graph is necessarily acyclic (no directed cycles are
formed) and allows the definition of event $r$'s depth, denoted by
$\mathit{depth}(r)$, as follows. Given a directed path in graph $P$ between two
events, let the path's length be defined as the number of edges on the path that
correspond to messages. That is, the edges contributing to the path's length are
those that fall under case (b) above in the definition of relation $B$. We
define $\mathit{depth}(r)$ as the length of the lengthiest path leading to $r$
in graph $P$. Intuitively, $\mathit{depth}(r)$ is the size of the longest causal
chain of messages leading up to event $r$ during run $R$. These notions are
illustrated in Figure~\ref{fig:graphs}.

\subsection*{Measures of synchronization}

We use two measures of synchronization. They are both based on the premise that,
if the sequence of events at two neurons (nodes of $S$) can be aligned with each
other so that sufficiently many event pairs of the same depth become matched in
the alignment, then there is more synchronism between the two sequences than
there would be with fewer event pairs aligned. We do this sequence alignment as
described next. Let $i$ and $j$ be the two neurons in question and let their
event sequences during run $R$ be $e_i^1,e_i^2,\ldots,e_i^{L_i}$ and
$e_j^1,e_j^2,\ldots,e_j^{L_j}$, respectively, where $L_i$ and $L_j$ are the
sequences' sizes. Let
$\mu_{ij}=\max\{\mathit{depth}(e_i^{L_i}),\mathit{depth}(e_j^{L_j})\}$, i.e.,
$\mu_{ij}$ is the depth of the last event of the two sequences that is deepest.
We do the alignment of the two sequences by creating two new size-$\mu_{ij}$
sequences, viz.\ $t_1,t_2,\ldots,t_{\mu_{ij}}$ and $u_1,u_2,\ldots,u_{\mu_{ij}}$
for the first measure, and $x_1,x_2,\ldots,x_{\mu_{ij}}$ and
$y_1,y_2,\ldots,y_{\mu_{ij}}$ for the second measure, each with defining
characteristics that depend on the particular measure of synchronization under
consideration.

The first measure aims to capture the synchronization that may exist in the
overall flow of messages and, through it, in the accumulation of potential at
the neurons' membranes. As such, it is based on positioning, relative to the $t$
sequence, those of the $L_i$ events of neuron $i$ that promoted the accumulation
of potential, and likewise the $L_j$ events of neuron $j$ relative to the $u$
sequence. For $k=1,2,\ldots,\mu_{ij}$, the $t$ sequence is defined recursively
as follows. If $k=1$:
\begin{itemize}
\item $t_k=1$, if $\mathit{depth}(e_i^1)=1$;
\item $t_k=0$, otherwise.
\end{itemize}
If $k>1$:
\begin{itemize}
\item $t_k=k$, if $\mathit{depth}(e_i^\ell)=k$ for some
$\ell\in\{1,2,\ldots,L_i\}$;
\item $t_k=t_{k-1}$, otherwise.
\end{itemize}
The $u$ sequence is defined entirely analogously for neuron $j$.

If, for example, the events in the two original sequences have depths
$2,3,3,7,8,9,11$ and $1,3,4,5,5,9$ for $i$ and $j$, respectively, then the $t$
sequence is
\begin{equation}
t=\langle 0,2,3,3,3,3,7,8,9,9,11\rangle
\end{equation}
and the $u$ sequence is
\begin{equation}
u=\langle 1,1,3,4,5,5,5,5,9,9,9\rangle.
\end{equation}
Thus, the $k$th position of sequence $t$ or $u$ equals $k'$ such that
$0<k'\le k$ if and only if the corresponding neuron received at least one
message of depth $k'$ during the run and, for $k'<k$, never since did the neuron
receive a message whose depth is in the interval $[k'+1,k]$. It equals $0$
otherwise.

The first measure of synchronization is denoted by $\rho_{ij}^-$ for neurons $i$
and $j$. It is given by
\begin{equation}
\rho_{ij}^-=\frac{1}{\mu_{ij}}
\sum_{k=1}^{\mu_{ij}}
\frac{\min\{t_k,u_k\}}{\max\{t_k,u_k\}}
\label{eq:rho-}
\end{equation}
(assuming $0/0\equiv 1$). Clearly, $\rho_{ij}^-\in[0,1]$ and grows with the
similarity of sequences $t$ and $u$. For identical sequences we get
$\rho_{ij}^-=1$. The example above yields
\begin{equation}
\rho_{ij}^-=
\frac{1}{11}
\left(
\frac{0}{1}+
\frac{1}{2}+
\frac{3}{3}+
\frac{3}{4}+
\frac{3}{5}+
\frac{3}{5}+
\frac{5}{7}+
\frac{5}{8}+
\frac{9}{9}+
\frac{9}{9}+
\frac{9}{11}
\right)
\approx 0.69.
\end{equation}

The second measure addresses the synchronization possibilities that may exist as
the neurons fire. In this case only those of the $L_i$ events of neuron $i$
that entailed firing are positioned relative to the $x$ sequence, and likewise
the $L_j$ events of neuron $j$ with respect to the $y$ sequence. For
$k=1,2,\ldots,\mu_{ij}$, the $x$ sequence is such that:
\begin{itemize}
\item $x_k=k$, if $\mathit{depth}(e_i^\ell)=k$ for some
$\ell\in\{1,2,\ldots,L_i\}$ such that neuron $i$ fired at the occurrence of
$e_i^\ell$;
\item $x_k=0$, otherwise.
\end{itemize}
The $y$ sequence is defined analogously for neuron $j$.

Following up on the examples given above, and assuming that neuron $i$ fired at
one of its depth-$3$ events and neuron $j$ fired at all its events except those
of depth $5$, the $x$ sequence is
\begin{equation}
x=\langle 0,0,3,0,0,0,0,0,0,0,0\rangle
\end{equation}
and the $y$ sequence is
\begin{equation}
y=\langle 1,0,3,4,0,0,0,0,9,0,0\rangle.
\end{equation}
Thus, the $k$th position of sequence $x$ or $y$ equals $k>0$ if and only if the
corresponding neuron fired upon receiving a depth-$k$ message. It equals $0$
otherwise.

For neurons $i$ and $j$, the second measure of synchronization is denoted by
$\rho_{ij}^+$ and given by
\begin{equation}
\rho_{ij}^+=\frac{1}{\mu_{ij}}
\sum_{k=1}^{\mu_{ij}}
\frac{\min\{x_k,y_k\}}{\max\{x_k,y_k\}}
\label{eq:rho+}
\end{equation}
(assuming $0/0\equiv 1$). As in the previous case, $\rho_{ij}^+\in[0,1]$ and
grows with the similarity of sequences $x$ and $y$. Identical sequences yield
$\rho_{ij}^+=1$. For the example above we get
\begin{equation}
\rho_{ij}^+=
\frac{1}{11}
\left(
\frac{0}{1}+
\frac{0}{0}+
\frac{3}{3}+
\frac{0}{4}+
\frac{0}{0}+
\frac{0}{0}+
\frac{0}{0}+
\frac{0}{0}+
\frac{0}{9}+
\frac{0}{0}+
\frac{0}{0}
\right)
\approx 0.73.
\end{equation}

For a fixed pair $i,j$ of neurons, both $\rho_{ij}^-$ and $\rho_{ij}^+$ seek to
characterize the possibility of synchronized behavior during run $R$. They do
this by quantifying the extent to which the events occurring at the two neurons
could be said to happen synchronously \textit{if a time basis existed common to
neurons $i$ and $j$ and time according to this basis elapsed along increasing
event depth.} Of course, our cortical model per se has no need for such a time
basis and is algorithmically correct regardless of any timing assumptions one
may make concerning the delay for action-potential propagation down the axons.
However, the kind of synchronized behavior we are investigating occurs at a
variety of temporal and spatial scales, therefore some time-related assumption
is inevitable if we are to extract any meaning out of the multitude of neuronal
events. Our choice seems reasonable because, conceptually, it contrasts with the
algorithm's inherent asynchronism only minimally: Every single pair of neurons
is given its own time basis for the computation of $\rho_{ij}^-$ and
$\rho_{ij}^+$ and this is reflected on the notationally explicit dependency of
$\mu_{ij}$ on the $i,j$ pair. Furthermore, viewing increasing event depth as
time may even have some biological plausibility to it. In fact, it appears that
the delay for an action potential to reach the various synapses connecting out
of the same axon is independent of how much of the axon actually has to be
traversed \cite{ilh94,sitk03}. In these terms, what our assumption does is to
generalize this independence for a group of axons.

Notwithstanding this common underlying feature of the two synchronization
measures, they are also markedly different in how they use event depth to assess
the similarity of two event sequences. In the case of $\rho_{ij}^+$, this is
done rather stringently, since only same-depth firing events and the $0/0$
ratios contribute to it. In the case of $\rho_{ij}^-$, these continue to be some
of the strongest contributors, but now they are joined by any pair of same-depth
events (not necessarily firing ones). Moreover, now an event's depth lingers
until a greater-depth event occurs and in the meantime continues to influence
$\rho_{ij}^-$.

\section*{Results}

Our computational results are based on the same methodology we followed in
\cite{nb11}, of which we now provide a brief review for the reader's benefit.
We use $v^0=-15$, $v^\mathrm{t}=0$, $\delta=0.0002$, and $\alpha=0.04$ in all
runs of algorithm $A$ on $S$. Each run starts at $m=50$ initiators chosen
uniformly at random and progresses until termination. A run is implemented as a
sequential program that, initially, selects an initiator randomly out of the $m$
that were chosen for the run, lets it fire, and queues up the messages it sends
for reception by the destination nodes. Subsequently, after all initiators have
had a chance to proceed in this way, a list is maintained containing all nodes
with nonempty input-message queues. One of them is selected at random and the
processing of its head-of-queue message is carried out (again with the possible
queuing of messages for consumption by other nodes). This is repeated until all
queues are empty.

The directed graph $S$ can be of one of types (i)--(iii), as explained next. In
order to keep the computational effort reasonably bounded with current
technology, the directed graph $D$ of which $S$ is the GSCC has $n=100$ nodes.
\begin{itemize}
\item[(i)] In this case $D$ is sampled from the cortical model described
above. As explained, the corresponding $S$ is expected to have $N\approx 90$
nodes. The expected in- or out-degree in $D$ is $3.7$.
\item[(ii)] In this case $D$ is sampled from the generalized
Erd\H{o}s-R\'{e}nyi model of directed graphs \cite{k90}. Given the expected in-
or out-degree $z$, an edge is placed from each node $i$ to each node $j\neq i$
with probability $z/(n-1)$. For consistency with type-(i) graphs we use $z=3.7$,
in which case $S$ is such that $N\approx 100$ with high probability.
\item[(iii)] In this case $D$ has a deterministic structure and is by
construction strongly connected. So $S=D$ and $N=100$. The structure of $D$ is
that of the directed circulant graph \cite{lpw01} with in- or out-degree equal
to $\lceil 3.7\rceil=4$. If we number the $n$ nodes $0,1,\ldots,n-1$, then in
$D$ every node $i$ has nodes $i+1$, $i+2$, $i+3$, and $i+4$ (modulo $n$) as
out-neighbors.
\end{itemize}
For each fixed $S$, $0.2N$ nodes are chosen uniformly at random to be
inhibitory, so long as no two of these nodes are connected by an edge [note
that, if $S$ is of type (iii), then all $20$ inhibitory nodes are in fact placed
deterministically, since necessarily they are found at equal intervals as we
traverse the nodes in the order $0,1,\ldots,n-1$]. Moreover, for each $S$ node
potentials and synaptic weights are chosen uniformly at random from the
intervals $[v^0,v^\mathrm{t}]$ and $[0,1]$, respectively. For type-(iii) graphs
this is the only source of nondeterminism.

We use $50$ $S$ instances of each type. For each fixed $S$ we use $50\,000$ run
sequences of algorithm $A$, each sequence comprising $10\,000$ runs. The first
run in a sequence starts from the node potentials and synaptic weights chosen
for the graph. Each subsequent run in the sequence starts from the node
potentials and synaptic weights at which the previous run ended. Along each
sequence we observe the behavior of $\rho_{ij}^-$ and $\rho_{ij}^+$ for each
pair $i,j$ of distinct nodes at six checkpoints. The first of these occurs
before any run actually takes place. Each of the remaining five occurs after
$2\,000$ additional runs have elapsed. The observation that takes place at a
checkpoint is based on $100$ additional runs, called side runs, each starting at
its own set of $m$ randomly chosen initiators and from the node potentials and
synaptic weights that are current at the checkpoint. At the end of the side
runs, the main sequence of runs is resumed from these same node potentials and
synaptic weights.

It is important to note that this computational setup would entail a
considerable amount of processing even if the side runs were excluded, since for
fixed $S$ algorithm $A$ would be run to completion $5\times 10^8$ times. These
runs are grouped into sequences so that the cumulative action of the model's
weight-update rule can be effected, but we also need many independent sequences
to account for the inherent nondeterminism of our model's asynchronous setting.
As defined, however, our synchronization measures are properties of a single
run, so they too need to be averaged out over many runs. We might have chosen to
do so directly over the runs that correspond to checkpoints, that is, over
$50\,000$ runs per checkpoint (one for each sequence). Each of these runs,
however, starts at the set of node potentials and synaptic weights that are
current in its sequence, so for each such set only one run would take place.
Introducing side runs has been a means to increase this number to $100$, with
the consequence of elevating the overall number of runs for fixed $S$ to
$5.3\times 10^8$.

The contribution of each side run concerning a fixed pair $i,j$ of distinct
nodes is to record the values of $\rho_{ij}^-$ and $\rho_{ij}^+$ at the end of
the run, as well as tag them with the labels 
$\delta_\mathrm{min}=\min\{\delta_{ij},\delta_{ji}\}$ and
$\delta_\mathrm{max}=\max\{\delta_{ij},\delta_{ji}\}$, where $\delta_{ij}$ and
$\delta_{ji}$ are the directed distances between the two nodes in $S$ (from $i$
to $j$ and from $j$ to $i$, respectively). After all $2.5\times10^8$ side runs
for a graph type have been completed at a checkpoint, we calculate the average
values of $\rho_{ij}^-$ and $\rho_{ij}^+$ over all node pairs having the same
tags. These averages are henceforth denoted by $\rho^-$ and $\rho^+$,
respectively.

\begin{figure}[p]
\begin{center}
\scalebox{0.950}{\includegraphics{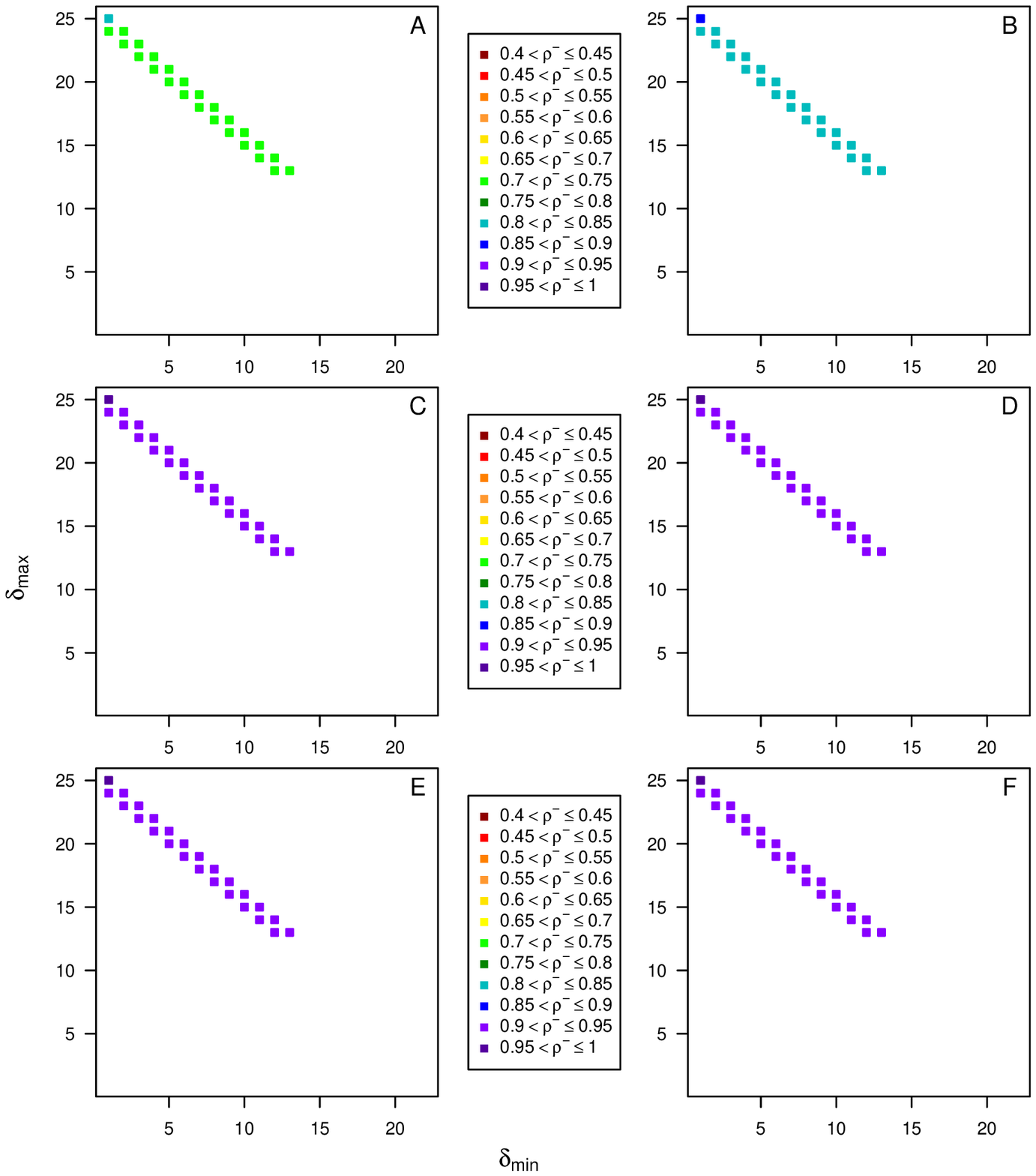}}
\end{center}
\caption{{\bf Average value of $\rho_{ij}^-$ for type-(iii) graphs as a function
of $\delta_\mathrm{min}$ and $\delta_\mathrm{max}$.} Data are given for the
first checkpoint (A) through the sixth (F).}
\label{fig:chain-ring}
\end{figure}

Our results are presented in Figures~\ref{fig:chain-ctx} and~\ref{fig:fire-ctx}
for type-(i) graphs, Figures~\ref{fig:chain-poisson} and~\ref{fig:fire-poisson}
for type-(ii) graphs, and Figures~\ref{fig:chain-ring} and~\ref{fig:fire-ring}
for type-(iii) graphs. The former figure in each pair refers to $\rho^-$, the
latter figure to $\rho^+$. Each figure comprises six panels, each panel for each
of the six observational checkpoints. The A panels refer to the first
checkpoints, the B panels to the second checkpoints, and so on. Each panel is
organized as a two-dimensional array and gives the $\rho^-$ or $\rho^+$ averages
as a function of the node pairs' $\delta_\mathrm{min}$ values (as abscissas) and
$\delta_\mathrm{max}$ values (as ordinates). We display these averages by means
of a color code that assigns different colors to different intervals inside
$[0,1]$ suitably. The hues we use vary from a dark shade of red to a dark shade
of purple, indicating the lowest interval and the highest one, respectively. We
note that this choice of colors is the same through all the figures and that the
colors always correspond to the same intervals.

Each average displayed in these figures refers to directed cycles in the $S$
graphs whose length is $\delta_\mathrm{min}+\delta_\mathrm{max}$ for the
particular $\delta_\mathrm{min}$ and $\delta_\mathrm{max}$ values in question.
Because of the strongly connected nature of $S$, every two nodes belong to at
least one common directed cycle. By the definitions of $\delta_\mathrm{min}$ and
$\delta_\mathrm{max}$, each average plotted in the figures is therefore relative
to all node pairs for which the shortest of these cycles has the same length. In
particular, traversing any of the array diagonals for which
$\delta_\mathrm{min}+\delta_\mathrm{max}$ is a constant merely shifts the
relative positions of the two nodes involved in each pair on the shortest
directed cycle that they share. We henceforth refer to the value of
$\delta_\mathrm{min}+\delta_\mathrm{max}$ for a certain node pair as the pairs'
girth.\footnote{In a free extension, to a node pair, of the homonymous notion in
graph theory that concerns the entire graph in the undirected case \cite{b98}.}
Moreover, we refer to the pair as being more or less balanced on the directed
cycle of length $\delta_\mathrm{min}+\delta_\mathrm{max}$, depending
respectively on whether $\delta_\mathrm{max}-\delta_\mathrm{min}$ is close to
$0$ or not (i.e., close to the $\delta_\mathrm{min}=\delta_\mathrm{max}$
diagonal of the array).

\begin{figure}[p]
\begin{center}
\scalebox{1.000}{\includegraphics{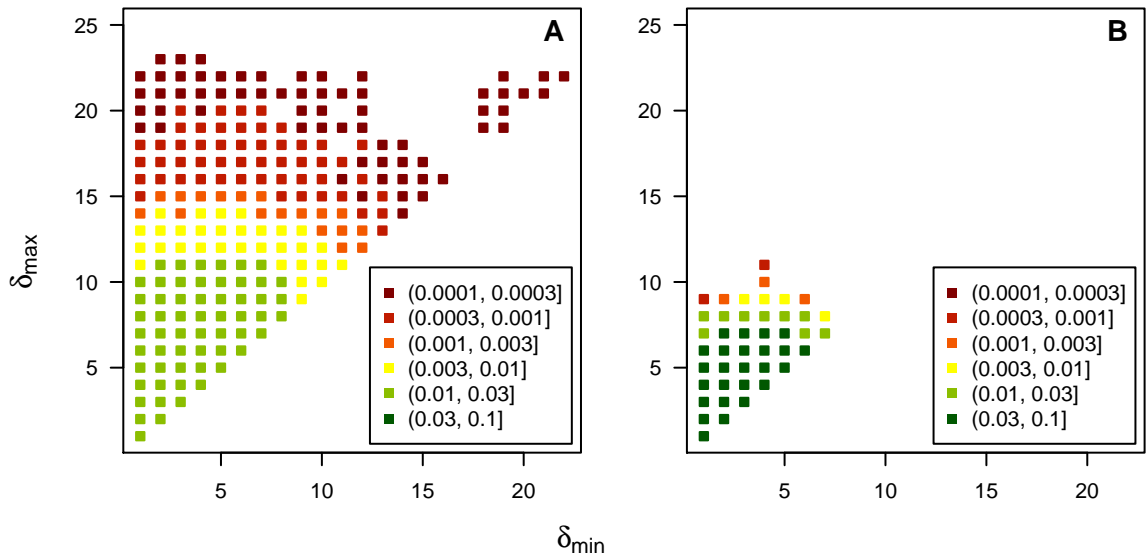}}
\end{center}
\caption{{\bf Probability distribution of
$\delta_\mathrm{min},\delta_\mathrm{max}$ pairs for the type-(i) (A) and
type-(ii) (B) graphs we used.}}
\label{fig:probabilities}
\end{figure}

All panels in Figures~\ref{fig:chain-ctx} through~\ref{fig:fire-ring} display
their data inside the upper triangular region relative to the
$\delta_\mathrm{min}=\delta_\mathrm{max}$ diagonal of the array. All blank spots
inside this region refer to $\delta_\mathrm{min},\delta_\mathrm{max}$ pairs that
never occurred in any of the $S$ instances we used. This can be verified by
resorting to Figure~\ref{fig:probabilities}, where the probability distributions
for the occurrence of these pairs are shown for type-(i) and type-(ii) graphs
(in panels A and B respectively, through the use of color codes similar to
those of the previous figures). As for type-(iii) graphs, it follows easily from
their definition that either $\delta_\mathrm{min}+\delta_\mathrm{max}=N/4$ or
$\delta_\mathrm{min}+\delta_\mathrm{max}=N/4+1$, respectively $25$ or $26$ for
$N=100$, so in this case these fixed girth values are the constraints
determining the appearance of blank spots (that is, they appear when the
constraints are violated).

Figure~\ref{fig:probabilities} is also useful in helping elucidate which of the
various possible girth values in type-(i) and type-(ii) graphs are the most
common. Readily, girths of about $18$ or less are by far the most common in
type-(i) graphs. This value becomes about $12$ for type-(ii) graphs. In the
forthcoming discussion, we use these rough delimiters to characterize what
happens to most node pairs (i.e., those whose girth values are overwhelmingly
the most common).

\section*{Discussion}

\begin{figure}[p]
\begin{center}
\scalebox{0.950}{\includegraphics{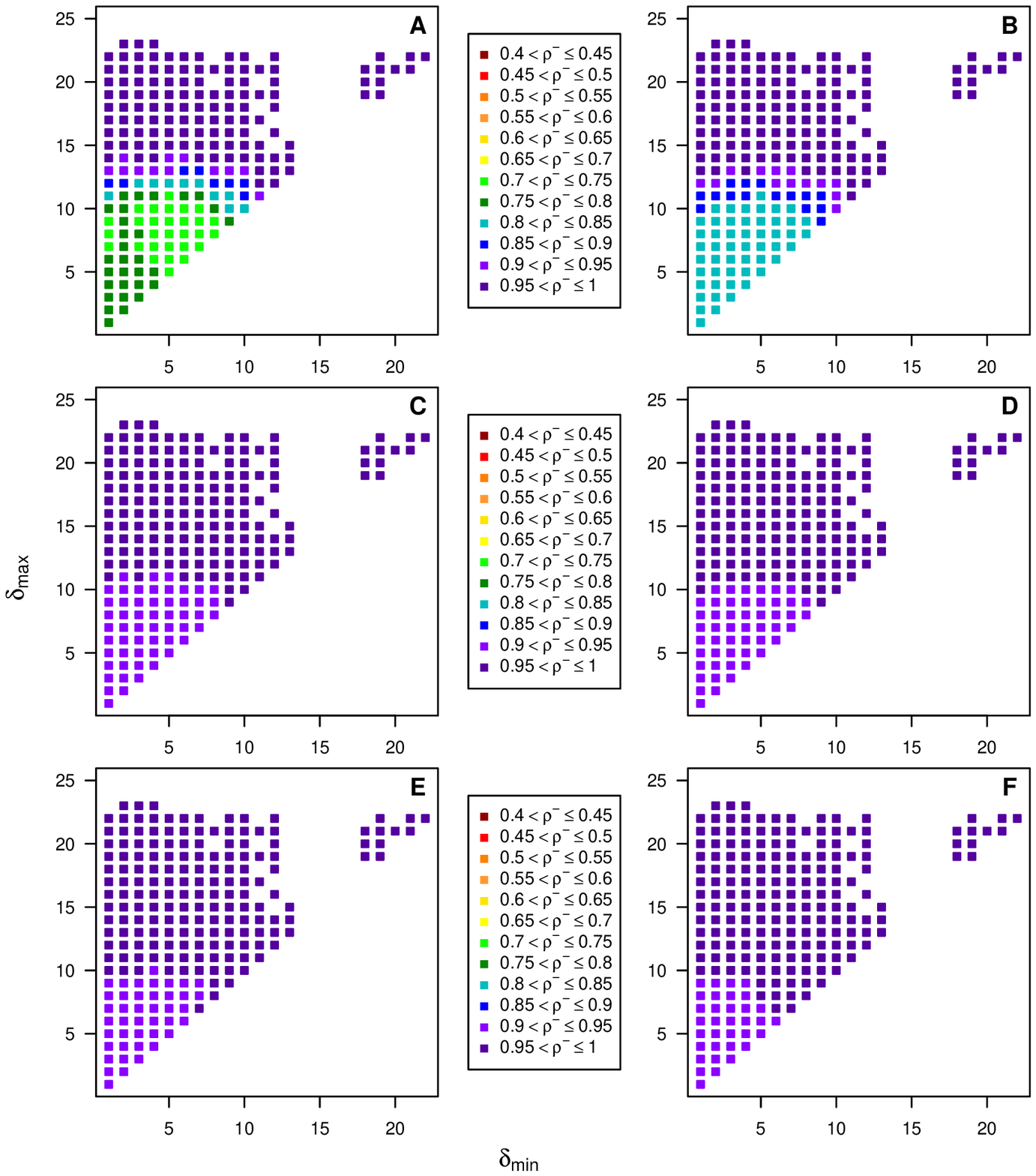}}
\end{center}
\caption{{\bf Average value of $\rho_{ij}^-$ for type-(i) graphs as a function
of $\delta_\mathrm{min}$ and $\delta_\mathrm{max}$.} Data are given for the
first checkpoint (A) through the sixth (F).}
\label{fig:chain-ctx}
\end{figure}

\begin{figure}[p]
\begin{center}
\scalebox{0.950}{\includegraphics{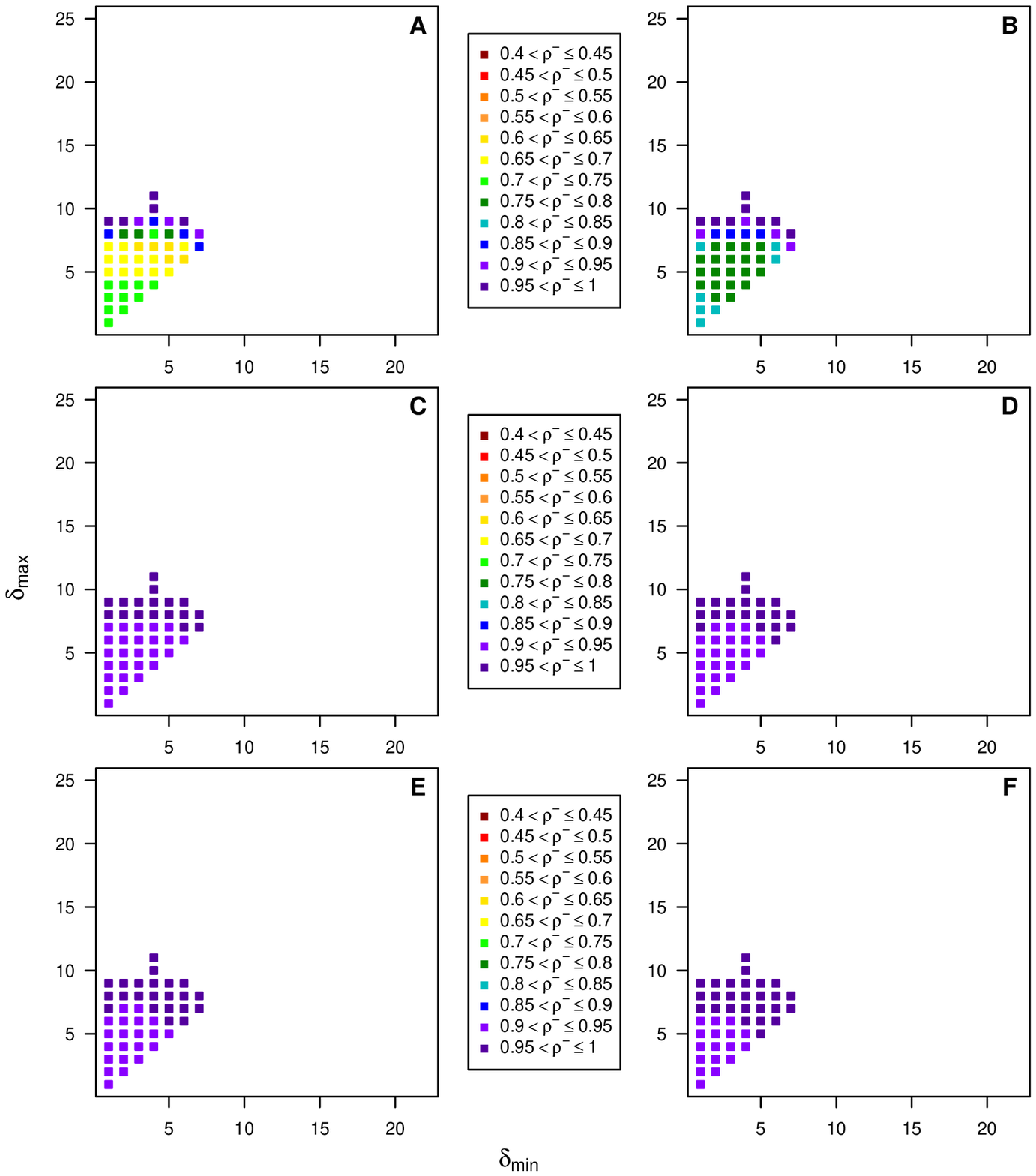}}
\end{center}
\caption{{\bf Average value of $\rho_{ij}^-$ for type-(ii) graphs as a function
of $\delta_\mathrm{min}$ and $\delta_\mathrm{max}$.} Data are given for the
first checkpoint (A) through the sixth (F).}
\label{fig:chain-poisson}
\end{figure}

The results for $\rho^-$, given in Figures~\ref{fig:chain-ctx},
\ref{fig:chain-poisson}, and~\ref{fig:chain-ring}, provide a clear picture of
what is to be expected regarding the overall synchronization that may be present
in the flow of messages as they get received at the neurons. In the case of
type-(i) graphs (Figure~\ref{fig:chain-ctx}), already at the first checkpoint
most node pairs $i,j$ have $\rho_{ij}^-$ values in the interval $(0.7,0.85]$.
Four thousand runs later (that is, at the third checkpoint) this holds for the
interval $(0.9,0.95]$. At the sixth and last checkpoint, the interval is
$(0.9,1]$. A closely analogous conclusion holds in the case of type-(ii) graphs
(Figure~\ref{fig:chain-poisson}), now with the intervals $(0.6,0.85]$ for the
first checkpoint, $(0.9,0.95]$ for the third, and $(0.9,1]$ for the last one.
As for type-(iii) graphs (Figure~\ref{fig:chain-ring}), their rigidly
constrained girth values lead $\rho_{ij}^-$ to be concentrated inside the
interval $(0.7,0.75]$ at the first checkpoint for nearly all node pairs.
Similarly, already at the third checkpoint the situation that we observe in the
last checkpoint has been established and $\rho_{ij}^-$ is concentrated in the
interval $(0.9,0.95]$ for nearly all node pairs.

It is curious to observe for type-(iii) graphs that, at all checkpoints, all
node pairs of girth $26$ for which $\delta_\mathrm{min}=1$ have $\rho_{ij}^-$
values occupying intervals one or two notches above the intervals we gave for
nearly all pairs, that is, $(0.8,0.85]$ for the first checkpoint and $(0.95,1]$
for the third checkpoint and onwards. Revisiting Figures~\ref{fig:chain-ctx}
and~\ref{fig:chain-poisson}, we see that a similar conclusion holds also for
type-(i) and type-(ii) graphs: Node pairs for which $\delta_\mathrm{min}$ is
near $1$ tend to have a slightly higher $\rho_{ij}^-$ value if their girth is
sufficiently large.

\begin{figure}[p]
\begin{center}
\scalebox{0.950}{\includegraphics{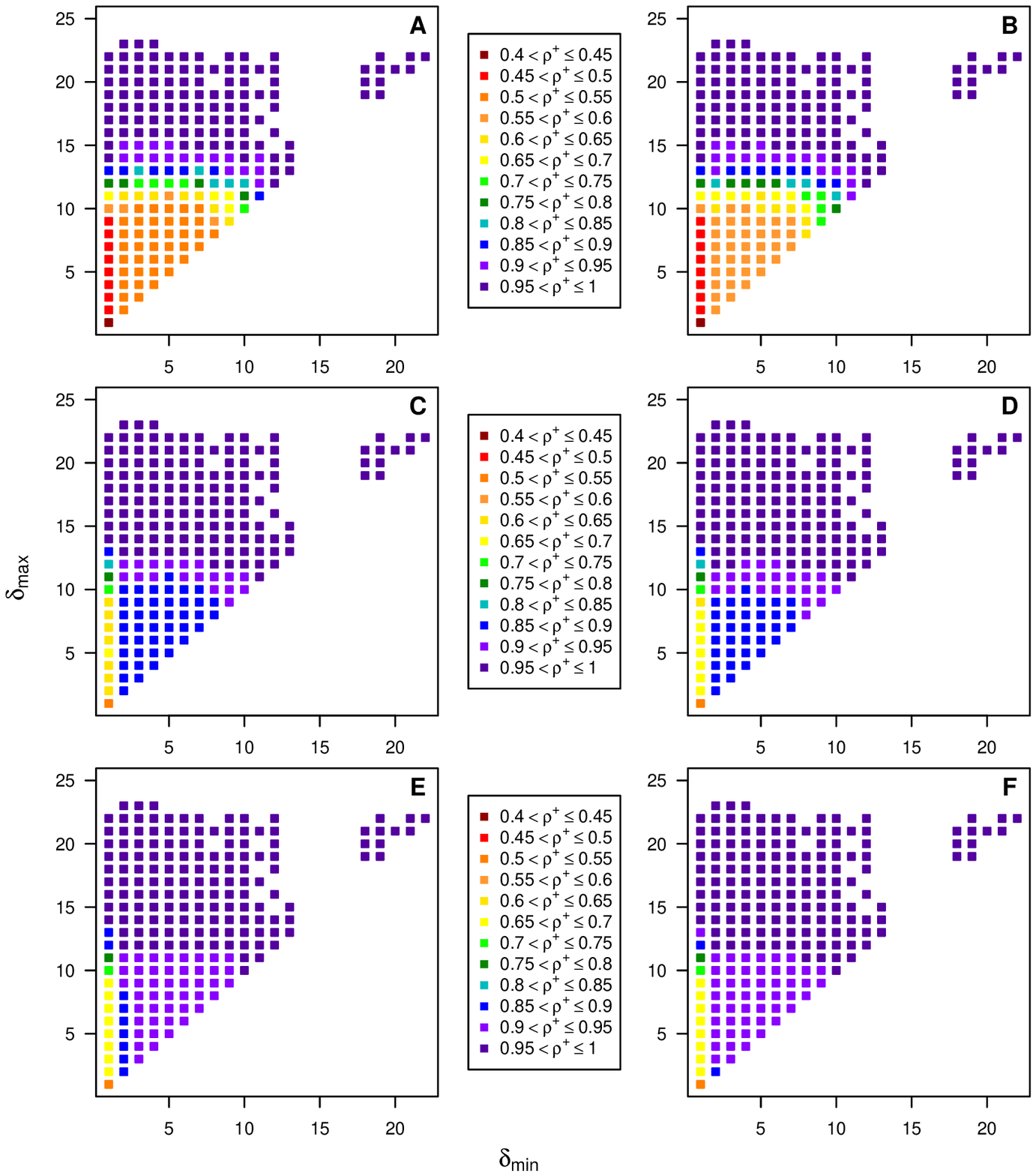}}
\end{center}
\caption{{\bf Average value of $\rho_{ij}^+$ for type-(i) graphs as a function
of $\delta_\mathrm{min}$ and $\delta_\mathrm{max}$.} Data are given for the
first checkpoint (A) through the sixth (F).}
\label{fig:fire-ctx}
\end{figure}

\begin{figure}[p]
\begin{center}
\scalebox{0.950}{\includegraphics{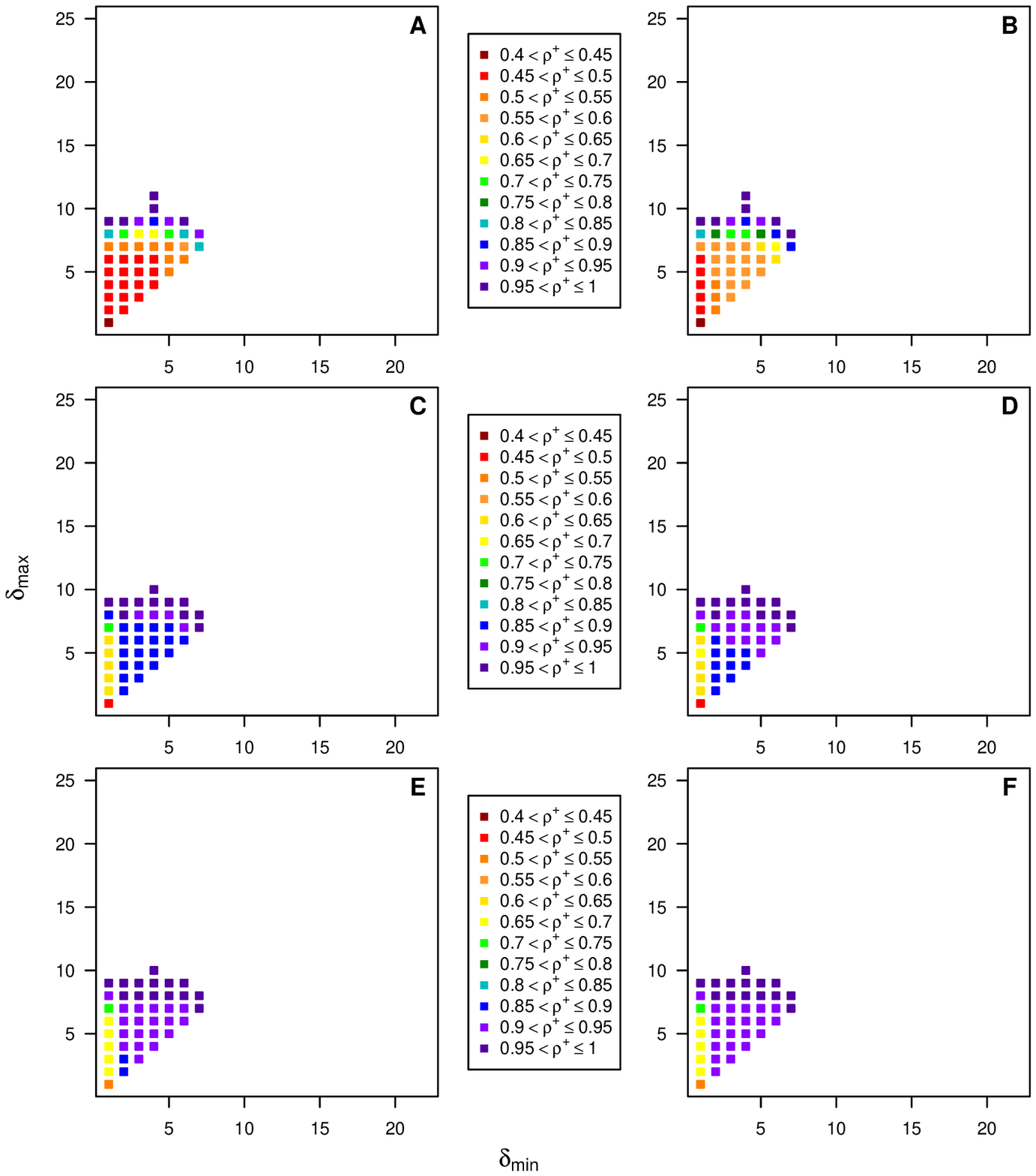}}
\end{center}
\caption{{\bf Average value of $\rho_{ij}^+$ for type-(ii) graphs as a function
of $\delta_\mathrm{min}$ and $\delta_\mathrm{max}$.} Data are given for the
first checkpoint (A) through the sixth (F).}
\label{fig:fire-poisson}
\end{figure}

\begin{figure}[p]
\begin{center}
\scalebox{0.950}{\includegraphics{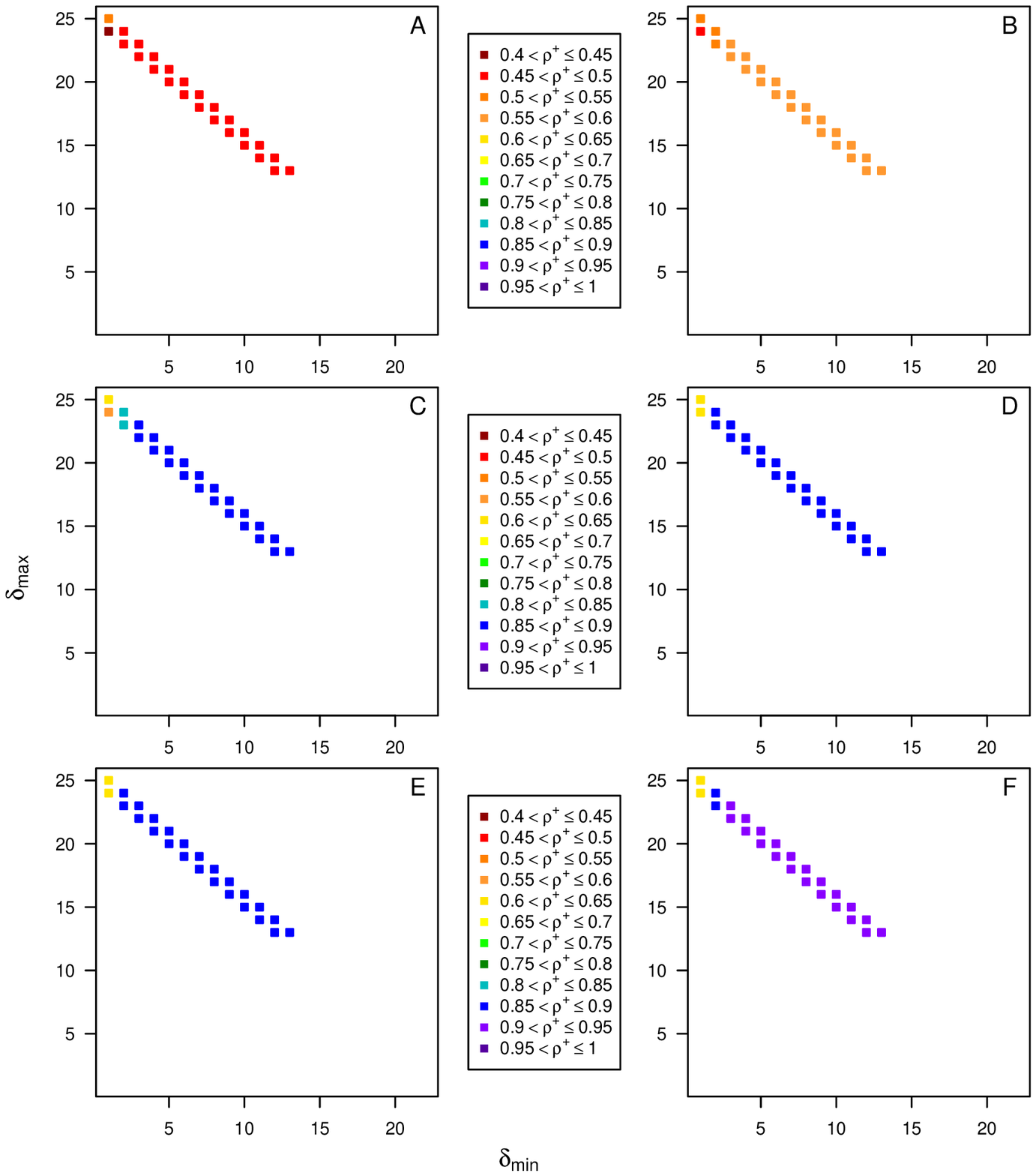}}
\end{center}
\caption{{\bf Average value of $\rho_{ij}^+$ for type-(iii) graphs as a function
of $\delta_\mathrm{min}$ and $\delta_\mathrm{max}$.} Data are given for the
first checkpoint (A) through the sixth (F).}
\label{fig:fire-ring}
\end{figure}

The results for $\rho^+$, which relates to how much synchronization may be
present as neurons fire, tell a story that differs from that of $\rho^-$ in
important ways. The first of these differences is clear from
Figures~\ref{fig:fire-ctx}, \ref{fig:fire-poisson}, and~\ref{fig:fire-ring}: For
all graph types the range of occurring $\rho^+$ values is wider by roughly
$50$--$100\%$ than that of the $\rho^-$ values. In fact, the $\rho^+$ values are
now scattered inside the $(0.4,1]$ interval for all graph types at the earliest
checkpoints and inside $(0.5,1]$ at the latest ones. Readily, then, according to
our two measures of synchronization there appear to be substantially fewer
synchronization possibilities in the firing of neurons than in the accumulation
of potential as reflected by the reception of messages. This can be quantified
by examining the data more closely, as follows.

For type-(i) graphs (Figure~\ref{fig:fire-ctx}), at the first checkpoint most
node pairs have $\rho_{ij}^+$ values in the interval $(0.4,0.8]$. At the last
checkpoint, if we ignore all node pairs for which $\delta_\mathrm{min}=1$ for
the time being then nearly all node pairs have $\rho_{ij}^+$ values in the
interval $(0.9,1]$ [the single exception is that of
$\delta_\mathrm{min}=\delta_\mathrm{max}=2$, for which the interval is
$(0.85,0.9]$]. The case of type-(ii) graphs (Figure~\ref{fig:fire-poisson}),
still disregarding all $\delta_\mathrm{min}=1$ entries, is closely analogous to
that of type-(i) graphs, the differences being that now most node pairs span the
larger interval $(0.4,1]$ already at the first checkpoint and that, without
exception, at the last checkpoint all node pairs hit the interval $(0.9,1]$.
Finally, if we go on focusing on node pairs for which $\delta_\mathrm{min}>1$
exclusively, then for type-(iii) graphs (Figure~\ref{fig:fire-ring}) all node
pairs have $\rho_{ij}^+$ values in the interval $(0.45,0.5]$ at the first
checkpoint. At the last checkpoint, on the other hand, this interval becomes
$(0.85,0.9]$ if $\delta_\mathrm{min}=2$, $(0.9,0.95]$ if
$\delta_\mathrm{min}>2$.

As Figures~\ref{fig:fire-ctx}, \ref{fig:fire-poisson}, and~\ref{fig:fire-ring}
demonstrate, the $\delta_\mathrm{min}=1$ case sets itself apart from the others
for all graph types at nearly all checkpoints. For example, if we concentrate on
the last checkpoint, at which we believe the model's dynamics to have already
settled into some sort of stationary regime \cite{nb10}, then for both type-(i)
and type-(ii) graphs it holds that $\rho_{ij}^+$ tends to increase from some
value in the interval $(0.55,0.6]$ when the girth is $2$ to some value in the
interval $(0.9,1]$ when the girth is close to the rough upper bound we set
earlier for declaring most node pairs to have been counted [i.e., roughly $18$
for type (i), roughly $12$ for type (ii)]. Naturally, increasing the girth while
$\delta_\mathrm{min}$ is held fixed at $1$ implies considering node pairs that
are progressively more imbalanced, since $\delta_\mathrm{max}$ grows with the
girth. The case of type-(iii) graphs is different in this respect, as for
$\delta_\mathrm{min}=1$ all node pairs have $\rho_{ij}^+$ values in the interval
$(0.6,0.65]$, regardless of which value of $\delta_\mathrm{max}$ is in question
(either $24$ or $25$). These pairs, however, are of course highly imbalanced as
well.

As we mentioned above, a notion that is becoming increasingly central to the
study of synchronization in the brain is that of feedback loops, both in a
temporal sense (as LFPs and neuronal firing patterns exert influence on one
another) and in a spatial sense (as the LFPs of spatially separated areas affect
one another). The results we have obtained with our algorithmic model of a
cortex lend support to this notion both in the temporal and in the spatial
sense.

In the temporal sense we have found ample evidence that our distributed
algorithm $A$ is capable of promoting abundant opportunities both for potential
to be accumulated in a synchronized way as messages are received at the neurons
and for neurons to fire in a synchronized manner. We have found that this holds
across all three graph types, from the cortical model first introduced in
\cite{nb10}, to an Erd\H{o}s-R\'{e}nyi directed graph, to the completely
deterministic and tautly shaped structure of a directed circulant graph. In our
view, this independence from the graph's structural characteristics points at an
inherent ability of algorithm $A$ at providing some of the elements that help
give rise to brain synchronization. Notwithstanding this, our results also do
shed some light on the role played by graph structure. As it happens, of the
nondeterministic graph types only type-(i) graphs provide the opportunity of
long-distance (in the sense of graph distances) synchronization in the two
senses we have studied, since distances in type-(ii) graphs are significantly
shorter.

In the spatial sense there are two important issues to be highlighted. The
first one is that, although by Figure~\ref{fig:probabilities} node pairs having
higher-than-$18$ girth are rare, they do occur and have yielded high $\rho^-$
and $\rho^+$ values at all the observational checkpoints. The exact significance
this may have in the case of real cortices is unknown, to the best of our
knowledge, since spatial feedback loops are known only for very small graph
distances \cite{f05,blt12}. So the fact that they may also occur at
significantly larger graph distances remains a tantalizing possibility. The
second important issue is the presence of such strong dependence of $\rho^+$ on
a node pair's girth when $\delta_\mathrm{min}=1$ as we observed. Our results
indicate that in this case the synchronization of neuronal spikes is favored on
feedback loops involving highly imbalanced pairs of neurons (i.e., node pairs
for which $\delta_\mathrm{max}\gg 1$). As with the first issue, the significance
this may have for real cortices is unknown and merits special attention as
further data are obtained.

All our results depend strongly on the measures of synchronization we gave in
Equations~(\ref{eq:rho-}) and~(\ref{eq:rho+}). They also depend on the model
summarized above, but that is now backed up by interesting validating finds
\cite{nb10,nb11} and has therefore proven its usefulness as an artificial-life
abstraction. It then seems that furthering our study of emerging synchronization
properties depends on validating our two measures in a way that ties our
causality-based definitions to real data as tightly as possible. We expect to be
able to do this as further insight into real cortices becomes available.

% Do NOT remove this, even if you are not including acknowledgments
\section*{Acknowledgments}

We acknowledge partial support from CNPq, CAPES, and a FAPERJ BBP grant.

%\section*{References}
% The bibtex filename
\bibliography{sync}

\clearpage
%\section*{Figure Legends}

\end{document}